\documentclass[aip,graphicx]{revtex4-1}
\usepackage{graphicx}
\usepackage{dcolumn}
\usepackage{bm}
\usepackage{float}
\usepackage{amsmath}
\usepackage{amsfonts}
\usepackage{amssymb}
\usepackage[utf8]{inputenc}
\usepackage[T1]{fontenc}
\usepackage{mathptmx}
\usepackage{etoolbox}
\usepackage{booktabs}
\draft 

\begin{document}


\title{Multifractal Signatures of Hamiltonian Chaos in Hyperion’s Rotational Dynamics} 



\author{S. Jaroszewicz}
\affiliation{Comisi\'on Nacional de Energ\'ia At\'omica, Bs. As., Argentina}

\author{N. Mendez}
\affiliation{Instituto Sábato. Universidad Nacional de San Martín Bs. As.,Argentina}

\author{Maria P. Beccar-Varela}
\affiliation{Department of Mathematical Sciences, UTEP. El Paso, United States}

\author{Maria Cristina Mariani}
\affiliation{Department of Mathematical Sciences, UTEP. El Paso, United States}


\date{\today}

\begin{abstract}
The chaotic rotation of Saturn’s moon Hyperion constitutes a canonical example of Hamiltonian chaos in a natural physical system. While its tumbling motion is well established theoretically, extracting a robust signature of chaos from sparse and noisy astronomical time series remains a fundamental challenge, rendering phase-space reconstruction methods largely impractical under realistic observational constraints. In this work, we demonstrate that multifractal detrended fluctuation analysis (MFDFA) provides an observationally viable alternative for identifying chaotic dynamics directly from photometric time series.

By analyzing historical ground-based light curves alongside synthetic data generated from the full three-dimensional Euler--Liouville equations, we show 
that the multifractality associated with the intermittency inherent to chaotic tumbling is consistent with a broad singularity spectrum. While multifractality is an established feature of Hamiltonian chaos, we show that it provides a novel and observationally viable diagnostic under sparse sampling—where traditional chaos indicators fail. In particular, we show that the multifractal spectrum survives realistic observational filtering and reliably discriminates intrinsic chaotic tumbling from aliased regular rotation in sparse astronomical time series. 
Conversely, regular resonant rotation exhibits a marked reduction in spectral width, approaching the monofractal limit characteristic of uncorrelated noise. 

For the observational data, we measure a broad spectral width that is consistent with the synthetic chaotic model, statistically distinct from surrogate datasets, and robust against the finite length of the observational time series. These results demonstrate the viability of multifractal scaling as a positive signature of Hamiltonian chaos in sparse datasets, bridging the gap between nonlinear dynamics theory and planetary photometry.
\end{abstract}


\maketitle 

\begin{quotation}
The identification of chaos in natural systems is often constrained by limited, irregular, or indirect observations, rendering conventional indicators such as Lyapunov exponents impractical. In this context, multifractal analysis has emerged as a powerful diagnostic tool for characterizing complex dynamical behavior directly from time series data. In this work, we apply multifractal detrended fluctuation analysis (MFDFA) to the rotational dynamics of Saturn’s moon Hyperion, a paradigmatic example of Hamiltonian chaos, and demonstrate that its observed and synthetic light curves exhibit robust multifractal signatures. By systematically validating the analysis, we establish multifractal spectra as reliable and interpretable fingerprints of chaotic dynamics in observationally constrained Hamiltonian systems.
\end{quotation}

\section{Introduction}
\subsection{Chaos in Celestial Rotation}
Hamiltonian chaos occupies a central position in modern nonlinear dynamics, representing the complex interplay between regular invariant tori and stochastic seas. Yet, direct empirical demonstrations of this fundamental behavior in natural systems remain rare. One of the most striking exceptions is the chaotic rotation of Saturn’s satellite Hyperion \cite{Binzel1986}. Unlike tidally evolved moons locked in synchronous rotation, Hyperion’s highly aspherical shape and eccentric orbit generate strong, time-dependent gravitational torques. These torques drive stochastic tumbling on timescales of weeks, a behavior predicted by Wisdom, Peale, and Mignard using the Chirikov resonance overlap criterion \cite{Wisdom1984}.

From the perspective of dynamical systems, Hyperion is a paradigmatic Hamiltonian system with a mixed phase space, dominated by overlapping spin–orbit resonances \cite{Goldberg2024}. However, from an observational standpoint, identifying this chaos is nontrivial. Astronomical time series are inherently finite, noisy, and sparsely sampled. These constraints severely limit the applicability of traditional metric diagnostics, such as Lyapunov exponents or phase-space reconstruction \cite{black1995hyperion, shevchenko2002chaotic}, which typically require dense, equidistant data to converge reliably.

\subsection{Limitations of Periodicity-Based Diagnostics and Spacecraft Observations}

Historically, the observational confirmation of Hyperion's chaotic state from ground-based photometry has relied primarily on negative evidence: the failure to identify a stable rotation period using techniques such as phase dispersion minimization~\cite{Klavetter1989}. While the absence of periodicity is consistent with chaos, it does not uniquely identify it. In sparse datasets, aliasing, undersampling, and stochastic noise can obscure periodic signals, producing time series that appear irregular even if the underlying system is deterministic and regular.

Beyond ground-based light curves, Hyperion's chaotic rotation has been robustly confirmed through the analysis of high-resolution imaging data from spacecraft flybys. Observations from Voyager 2~\cite{Thomas1995} and Cassini~\cite{Thomas2007, Harbison2011} provided detailed shape models and principal body axis ratios. These models allowed for accurate short-term numerical integrations of Hyperion's rotational motion, confirming an exponential divergence with a Lyapunov time of $60-100$ days~\cite{Thomas1995, Harbison2011}. Furthermore, data from the Cassini close flybys in 2005 revealed an instantaneous spin rate of $4.2-4.5n$ (where $n$ is the orbital mean motion) and a spin vector within $\sim 40^\circ$ of the longest body axis (the $a$-axis), states that are incompatible with both a regular spin state and a homogeneous interior~\cite{Harbison2011}. More recently, Zubarev and Nadezhdina (2025) reanalyzed Cassini images to generate an improved digital shape model and refined estimates of the principal moments of inertia, further constraining the satellite's physical parameters~\cite{Zubarev2025}.

However, while spacecraft imaging provides invaluable snapshots and confirms chaotic divergence via short-term integration, long-term continuous monitoring relies heavily on ground-based photometric time series. A robust diagnostic of chaos for these sparse series must therefore satisfy three criteria: (i) it must be resilient to sparse sampling and observational noise; (ii) it must distinguish low-dimensional deterministic chaos from high-dimensional stochastic processes; and (iii) it must encode information about the underlying phase-space structure. The limitations of standard periodicity searches underscore the need for a diagnostic that probes the dynamical structure of the fluctuations rather than merely the presence or absence of coherent frequencies.

\subsection{Multifractality and Hamiltonian Chaos}

Chaotic Hamiltonian systems are characterized by the nonuniform exploration of phase space. Unlike dissipative systems where trajectories contract onto a strange attractor, Hamiltonian dynamics preserve phase-space volume. The resulting mixed phase space contains chaotic layers interwoven with hierarchical chains of stability islands and remnants of invariant tori, as described by Kolmogorov–Arnold–Moser (KAM) theory \cite{goldberg2024nutation}.

A defining feature of Hamiltonian systems with mixed phase space is the presence of intermittent transport induced by stickiness near invariant tori, cantori, and hierarchical island chains. Chaotic trajectories may become temporarily trapped near these structures, leading to power-law-distributed trapping times, algebraic decay of correlations, and anomalous transport properties \cite{Meiss,Zaslavsky2002,Chirikov_1999}.  A clear illustration of this ``sticky'' behavior in Hyperion's specific case is found in the long-term numerical integrations by Black et al. (1995)~\cite{Black1995}. Notably, these pseudo-regular trapping episodes are frequently characterized by a spin rate of $\omega \approx 4n$ and rotation predominantly about the longest principal axis (the A-axis), closely mirroring the instantaneous states observed during the 1981 Voyager 2 and 2005--2007 Cassini spacecraft flybys~\cite{Black1995, Harbison2011}. Such intermittent dynamics have long been recognized as a generic mechanism in Hamiltonian chaos and provide the physical basis for the emergence of multifractal scaling in time series observables~\cite{Tarquis,osborne1991hamiltonian}.

When projected onto a scalar observable---such as the photometric light curve of a tumbling body---this phase-space intermittency manifests as a hierarchy of scale-dependent fluctuations. This phenomenon, explicitly known as \textit{multifractal scaling}, occurs when a time series cannot be adequately characterized by a single global scaling exponent (monofractality). Instead, it requires a continuous spectrum of local scaling exponents to describe its structural complexity, as different segments of the trajectory scale differently depending on whether the system is in a sticky, pseudo-regular regime or a strongly chaotic one. \cite{tarquis2001multifractal}.

Multifractal detrended fluctuation analysis (MFDFA) provides a natural and statistically robust framework for quantifying this hierarchy in finite and nonstationary time series \cite{Kantelhardt2002,kantelhardt2001,Ihlen2012}. It is a technique that has proven its capabilities even with complex data sets such as climatological data\cite{JAROSZEWICZ2024106161,Maruyama2011}, biological data such as DNA data \cite{Peng1994,Jaroszewicz2022}, volcanic measurments \cite{Mariani2020a,Mariani2020b} and astronomical data such as gravitational waves \cite{chernogor2017multi,de2018multifractal,cavaglia2022characterization}, among others. its application to planetary rotation offers a direct link to the underlying Hamiltonian dynamics.

Rather than relying on phase-space reconstruction or exponential divergence rates, MFDFA characterizes the heterogeneous scaling structure generated by intermittent Hamiltonian dynamics directly in the time domain. In contrast, regular or quasi-periodic Hamiltonian motion is supported on smooth invariant tori and is associated with bounded, nearly periodic fluctuations.  We hypothesize that the "stickiness" of the Hamiltonian phase space generates a broad multifractal singularity spectrum, whereas regular rotation (even when aliased) collapses to a narrow, near-monofractal limit.

\subsection{Limitations of Standard Chaos Diagnostics}

Given the intermittent dynamics and scale-free transport mechanisms described above, chaos diagnostics that rely on dense phase-space reconstruction become particularly fragile when applied to finite and irregularly sampled astronomical time series. A wide range of diagnostics has been developed to detect chaos, including Lyapunov exponents \cite{Wolf1985,Arnold1963}, recurrence-based methods \cite{Marwan2007}, and the 0--1 test for chaos \cite{GottwaldMelbourne2004}. While powerful under controlled conditions, these techniques rely on assumptions that are frequently violated in ground-based planetary photometry.

While continuous numerical integrations of Hyperion's rotational dynamics demonstrate clear exponential divergence with Lyapunov times on the order of $60$ to $100$ days~\cite{Black1995}, the estimation of Lyapunov exponents from scalar time series typically requires reliable phase-space reconstruction through delay embedding. In astronomical settings, the limited duration of observations, uneven sampling, and noise contamination strongly undermine the stability of embedding-based methods. Thus, it is the severe observational filter---rather than the intrinsic stickiness of the Hamiltonian phase space---that typically obscures the measurement of exponential divergence over finite observational windows.

Similarly, Recurrence Plots (RP) and Recurrence Quantification Analysis (RQA) provide a geometry-based perspective on dynamical complexity but depend sensitively on threshold choices and embedding parameters. For sparsely sampled or irregularly spaced data, recurrence structures can be dominated by sampling artifacts rather than intrinsic dynamics, complicating the statistically robust identification of chaos. The 0--1 test for chaos offers a conceptually simple binary classification but similarly requires long, uniformly sampled time series to achieve reliable convergence. In the presence of observational gaps or significant noise, the test often yields inconclusive intermediate values, limiting its applicability to real astronomical data.

In contrast, MFDFA operates directly on the time series, avoiding the need for explicit phase-space reconstruction. By quantifying scale-dependent fluctuations, it provides a characterization of dynamical complexity that is robust to nonstationarity and---when coupled with appropriate interpolation---sparse sampling. Rather than producing a binary classification, the resulting multifractal spectrum captures the hierarchical structure of fluctuations associated with the underlying Hamiltonian intermittency. As we will demonstrate in Section \ref{sec:results} (see explicitly Fig. \ref{fig:spectrum_comparison}), this multifractal signature remains remarkably robust and diagnostically powerful even when the underlying data is subjected to severe observational constraints.

\subsection{Scope and Structure of This Work}
In this paper, we show multifractality as a positive and observationally resilient signature of Hamiltonian chaos. We combine historical photometric observations of Hyperion with synthetic light curves generated from full three-dimensional rigid-body dynamics. Through systematic robustness tests and surrogate data analysis (shuffling and IAAFT), we demonstrate that the multifractal properties observed in the real data are intrinsic to the dynamics and cannot be explained by linear correlations, noise, or finite-size effects.

We emphasize that multifractality in Hamiltonian systems with mixed phase space is not a newly discovered phenomenon. Its origin in intermittent transport, 
stickiness near invariant structures, and algebraic decay of correlations has been extensively discussed in the nonlinear dynamics literature. 
The contribution of the present work is therefore not the identification of multifractality per se, but the demonstration that this scaling hierarchy 
constitutes a practically extractable and statistically robust diagnostic of Hamiltonian chaos under realistic observational constraints.

In this sense, our study addresses a fundamentally operational question: whether signatures of mixed phase-space dynamics, known from theory and 
high-resolution simulations, can be reliably detected in sparse, noisy, and  irregularly sampled astronomical time series where conventional chaos 
diagnostics are infeasible.

Our results position the multifractal spectrum width as a powerful diagnostic tool for identifying chaotic dynamics in sparse astronomical time series. This framework holds broad applicability to tumbling non-spherical moons, non-principal axis rotating (``tumbling'') asteroids~\cite{Pravec2005}, and other rotationally complex bodies. We explicitly note that in planetary astronomy, the term ``irregular satellites'' strictly defines outer moons with highly eccentric and significantly inclined orbits~\cite{Nicholson2008}, whose rotational states are not necessarily irregular. Here, our focus is exclusively on bodies exhibiting complex or chaotic \textit{rotational} dynamics, independent of their orbital classification. 

The article is structured as follows: Section II details the numerical integration of the full three-dimensional Euler-Liouville equations used to generate synthetic light curves , alongside the cubic spline regularization technique employed to handle sparse observational data. This section also outlines the implementation of the Multifractal Detrended Fluctuation Analysis (MFDFA) formalism. Section III presents the comparative analysis of chaotic and regular rotational states, quantifying their multifractal signatures under realistic observational filters and validating the results through extensive surrogate data and robustness testing. Section IV discusses the physical origin of the observed multifractality, linking it to the intermittent stickiness characteristic of Hamiltonian mixed phase space , and addresses the scope and limitations of the method. Finally, Section V summarizes our findings and their implications for future observations of natural satellites exhibiting irregular shapes and complex rotational behavior.

\section{Methods}
\subsection{Dynamical Model and Synthetic Light Curves}
To provide a rigorous control dataset for our multifractal analysis, we generated synthetic light curves based on the Hamiltonian dynamics of a rigid, triaxial satellite. Unlike simplified planar models, our simulation accounts for the full three-dimensional tumbling motion, which is essential for accurately modeling the projected cross-sectional area observed from Earth.

We model Hyperion as a triaxial ellipsoid with principal semi-axes $a > b > c$. While the assumption of a homogeneous ellipsoid is a standard and necessary mathematical idealization for constructing a baseline synthetic dynamical model, we acknowledge that this is a simplification. As noted by Harbison et al. (2011), Cassini observations of Hyperion's spin axis orientation, combined with its extremely low macroscopic density, strongly suggest a highly porous and non-homogeneous interior~\cite{Harbison2011}. 

Following the dynamical framework established by Tarnopolski (2015), we adopt the dimensions $a=180$~km, $b=140$~km, and $c=112$~km for the principal semi-axes~\cite{Tarnopolski2015}. The principal moments of inertia $A < B < C$ correspond to rotation about the axes $a$, $b$, and $c$, respectively. To ensure the reproducibility of our numerical results, the specific initial conditions for both the regular and chaotic rotational states---including the initial Euler angles and the initial body-frame angular velocity components---were adopted exactly as detailed in the numerical setup by Tarnopolski (2015)~\cite{Tarnopolski2015}.
The complete set of physical and orbital parameters used in the simulation is listed in Table~\ref{tab:params}.

\begin{table}[h!]
\caption{\label{tab:params} Physical and orbital parameters of the Saturn-Hyperion system used in the numerical simulation. The physical dimensions correspond to a homogeneous triaxial ellipsoid model.}
\begin{ruledtabular}
\begin{tabular}{lccc}
\textbf{Parameter} & \textbf{Symbol} & \textbf{Value} & \textbf{Unit} \\
\hline
Saturn's Mass\footnote{Standard value assumed.} & $M$ & $5.68 \times 10^{26}$ & kg \\
Orbital semi-major axis\textsuperscript{a} & $a_{orb}$ & $1.43 \times 10^{6}$ & km \\
Orbital eccentricity & $e$ & $0.1042$ & - \\
Orbital period\textsuperscript{a} & $T_{orb}$ & $21.28$ & d \\
Principal semi-axes & $a \times b \times c$ & $180 \times 140 \times 112$ & km \\
Inertial parameter\footnote{Derived from $a,b,c$ for a homogeneous ellipsoid: $\omega_0^2 = 3(B-A)/C$.} & $\omega_0^2$ & $0.74$ & - \\
\end{tabular}
\end{ruledtabular}
\end{table}

The rotational state is defined by the Euler angles $(\phi, \theta, \psi)$ and the body-frame angular velocity vector $\vec{\omega}$. The motion is governed by Euler's equations for a rigid body subject to external gravitational torque $\vec{N}$ from Saturn:

\begin{equation}
\begin{aligned}
A \dot{\omega}_x - (B - C) \omega_y \omega_z &= N_x \\
B \dot{\omega}_y - (C - A) \omega_z \omega_x &= N_y \\
C \dot{\omega}_z - (A - B) \omega_x \omega_y &= N_z
\end{aligned}
\end{equation}

In the body frame, if $\vec{u}$ is the unit vector pointing toward Saturn, the torque components are:
\begin{equation}
N_x = \frac{3GM}{r^3} (C - B) u_y u_z, \quad N_y = \frac{3GM}{r^3} (A - C) u_z u_x, \quad N_z = \frac{3GM}{r^3} (B - A) u_x u_y
\end{equation}

where $r(t)$ is the instantaneous orbital distance.

The orbital motion is treated as a fixed Keplerian ellipse. It is important to note that Hyperion's relatively high orbital eccentricity ($e \approx 0.10$)---a critical ingredient for its chaotic rotation alongside its aspherical shape---is driven and maintained by a 3:4 mean-motion resonance with the much larger satellite Titan. The approximation of a fixed orbit is justified by the strong separation of timescales between the fast chaotic rotation (days) and the secular orbital evolution (years), allowing us to treat orbital variations as adiabatic perturbations that do not qualitatively affect the phase-space structure or the statistical properties of the rotational chaos investigated here.

We integrated the coupled system of 3D differential equations using a high-order adaptive Runge-Kutta scheme with strict error tolerances ($10^{-10}$) to ensure energy stability over 100 orbital periods. Initial conditions were selected to place the satellite in a chaotic zone. Specifically, the chaotic state was initialized slightly off the 2:1 spin-orbit resonance (with an initial spin rate of $\omega \approx 2.1n$) and with a non-zero obliquity to trigger full 3D tumbling. We verified through our numerical integrations that the resulting trajectory successfully explores the extended chaotic sea; the instantaneous spin rate frequently reaches and exceeds $\omega \approx 4n$, which is entirely consistent with the high-spin dynamical states observed during the Voyager 2 and Cassini flybys.

\paragraph*{Observable Calculation and Geometry.}
The key observable is the integrated flux, which depends on the geometric cross-section. The projected area $S(t)$ of the ellipsoid seen from a viewing unit vector $\hat{O}$ is given by:
\begin{equation}
S(t)=\pi\sqrt{b^2 c^2(\hat{O}\cdot\hat{x}_{b})^{2}+c^2 a^2(\hat{O}\cdot\hat{y}_{b})^{2}+a^2 b^2(\hat{O}\cdot\hat{z}_{b})^{2}}
\end{equation}
where $\hat{x}_{b}$, $\hat{y}_{b}$, and $\hat{z}_{b}$ are the body's principal axes in the inertial frame. The synthetic magnitude is derived as $m_{syn}(t)=-2.5 \log_{10}(S(t))$.

We employ a fixed-observer approximation ($\hat{O}=[1,0,0]$). It is important to note that the observed magnitude is strictly proportional to $S(t)$ only at zero phase angle, when the entire disk is illuminated. However, since the maximum phase angle of Hyperion as observed from Earth is only $\sim 6^\circ$, this fixed-observer, zero-phase-angle assumption is a highly reliable approximation. Furthermore, any slow geometric modulations in the phase angle due to the orbital period and Earth-Saturn synodic cycle appear as low-frequency trends. By employing second-order detrending (DFA2) in our multifractal analysis, these macroscopic trends are locally removed and do not contaminate the high-frequency scaling behavior of the chaotic fluctuations.

\subsection{Data Preprocessing: Grid Regularization}
Standard MFDFA algorithms require time series indexed on a uniform grid. However, ground-based astronomical observations are inherently irregular. To address this, we adopt the regularization protocol established by Tarnopolski \cite{Tarnopolski2015}.

We fit a natural cubic spline to the irregular magnitude observations and resample the continuous function onto a uniform grid $T_j$. Crucially, to prevent the introduction of spurious high-frequency artifacts, the resampling cadence is set strictly to the \textit{mean} sampling interval of the original dataset ($\Delta t_{grid} = \langle \Delta t_{obs} \rangle$). This ensures that the spline reconstruction preserves the power spectral density (PSD) structure of the original signal without "inventing" data at scales finer than the observational resolution.

\begin{figure}[htbp]
    \centering
    \includegraphics[width=\linewidth]{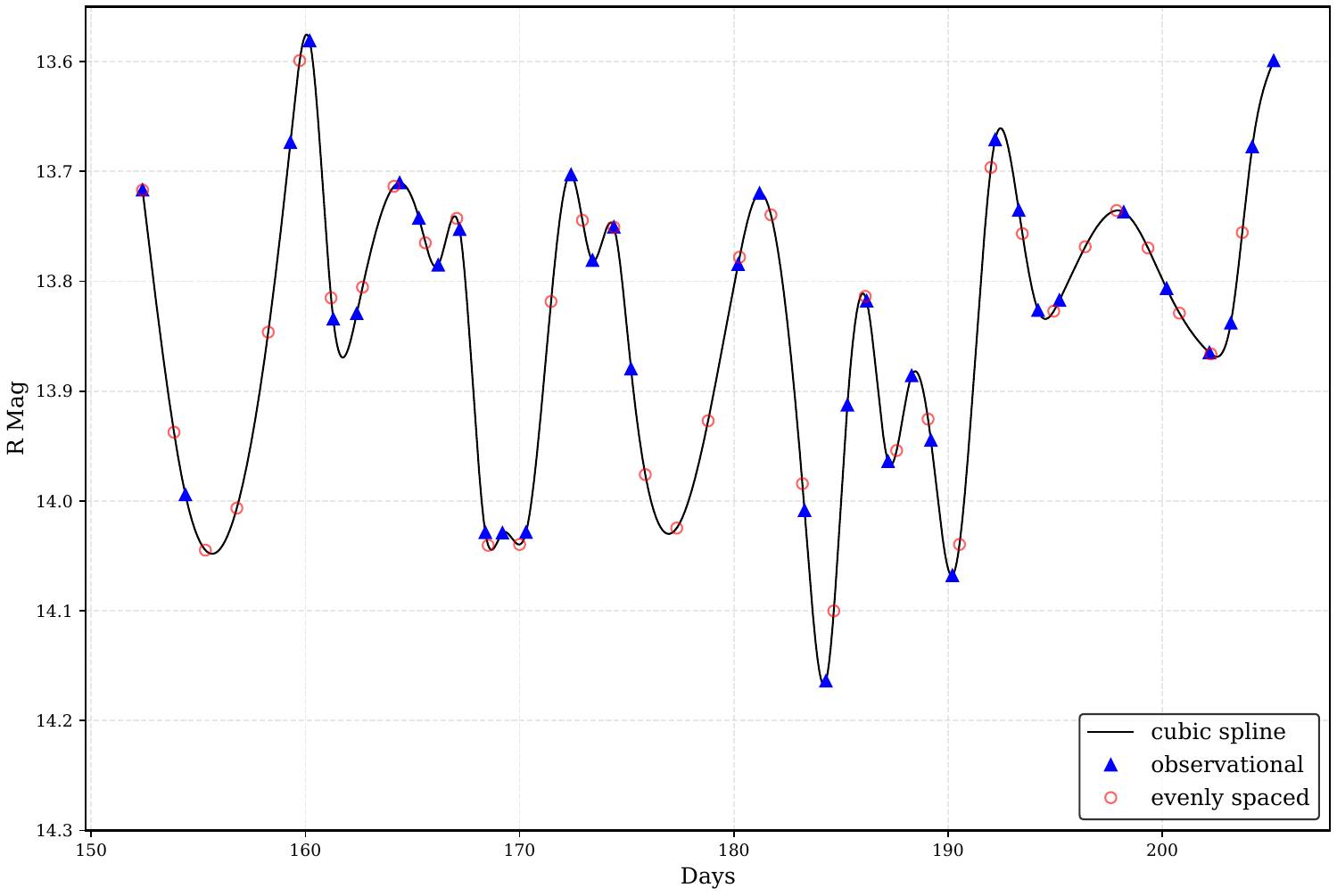}
    \caption{\label{fig:reconstruction}
    \textbf{Regularization of the Photometric Time Series.}
    Comparison between the irregular ground-based observations from the Klavetter (1989)~\cite{klavetter} dataset (blue triangles) and the natural cubic spline reconstruction (black solid line) used for MFDFA. Note that this figure displays a representative 50-day sample extracted from the full dataset (specifically from Table III of Klavetter, 1989~\cite{klavetter}) to clearly illustrate the interpolation behavior.
    The spline is resampled at the mean observational cadence ($\Delta t \approx 1.0$ day).
    The interpolation strictly respects the observational tie-points while bridging the sampling gaps, thereby preserving the macroscopic variability of the light curve without introducing high-frequency artifacts between epochs.
    }
\end{figure}

\subsection{Noise Modeling: Robustness Against Noise Color}

Real observations are contaminated by noise that is often non-Gaussian. To simulate realistic conditions, we contaminated our synthetic light curves with additive noise. Beyond simple Gaussian white noise ($\beta=0$), we also generated ``colored'' noise profiles (Red/Pink noise, $S(f)\propto 1/f^\beta$) to simulate atmospheric transparency variations. 

We observed that while strongly correlated noise ($\beta \ge 1.5$) shifts the crossover scale---the specific temporal scale at which the dominant fluctuation behavior changes---to larger values, the macroscopic multifractal signature remains intact in the scaling range $s \in [10, 100]$ days, where $s$ denotes the time-window size used in the fluctuation analysis. Since colored noise is inherently monofractal, it yields a constant generalized Hurst exponent $h(q)$ across different moment orders $q$. Consequently, the addition of colored noise does not artificially generate the $q$-dependent curvature in $h(q)$ that is the hallmark of multifractal chaotic dynamics (the formal mathematical definitions of $s$ and $h(q)$ are introduced in Section II.D, with their visual representations discussed alongside Figure ~\ref{fig:spectrum_comparison}).

\subsection{Multifractal Detrended Fluctuation Analysis (MFDFA)}

To quantify the scaling properties and dynamical complexity of the light curves, we employ the Multifractal Detrended Fluctuation Analysis (MFDFA) formalism \cite{Kantelhardt2002}. Unlike standard autocorrelation functions or power spectral density analyses, which quantify linear correlations globally, MFDFA reveals the hierarchical structure of fluctuations by examining scaling behaviors across multiple timescales and fluctuation magnitudes. This is particularly relevant for Hamiltonian systems, where the mixed phase space leads to intermittent dynamics that cannot be fully described by a single scaling exponent. The analysis proceeds in five distinct steps:

\textbf{Step 1: Profile Construction.} We first integrate the mean-adjusted time series $x_k$ (where $x_k$ represents the synthetic or observed magnitudes) to create the cumulative profile:
\begin{equation}
    Y(i) = \sum_{k=1}^{i} [x_k - \langle x \rangle]
\end{equation}
While the observational data are logarithmic fluxes (magnitudes), the cumulative sum $Y(i)$ does not represent a physical product or accumulation of fluxes. Rather, it is a standard mathematical transformation that maps a bounded, stationary-like noise signal into a non-stationary random walk profile. This transformation is necessary to robustly assess variance scaling without being bounded by the finite amplitude of the raw data.

\textbf{Step 2: Segmentation.} The profile $Y(i)$ is divided into $N_s \equiv \text{int}(N/s)$ non-overlapping segments of equal length $s$, where $s$ is the time-window size. Because the total length $N$ is rarely a multiple of $s$, a short part at the end of the profile may be discarded. To ensure no data from the tail is lost, the segmentation is repeated starting from the opposite end of the series. This yields a total of $2N_s$ segments. This dual-pass approach does not imply the dataset is treated as periodic; it simply maximizes data utilization.

\textbf{Step 3: Local Detrending.} For each of the $2N_s$ segments, we calculate the local trend by fitting a polynomial $y_\nu(i)$ of order $m$ (e.g., $m=2$ for DFA2). We then determine the local variance, $F^2(\nu, s)$, which represents the local mean squared error (or variance) of the profile $Y$ relative to the polynomial fit:
\begin{equation}
    F^2(\nu, s) = \frac{1}{s} \sum_{i=1}^{s} \{ Y[(\nu-1)s + i] - y_\nu(i) \}^2
    \label{eq:dfa}
\end{equation}
For $m=2$, the square root of this value, $F_2(s)$, is simply the local root-mean-square (RMS) deviation from the quadratic model.

\textbf{Step 4: Fluctuation Function.} We average the local variances across all segments to obtain the $q$-th order fluctuation function:
\begin{equation}
    F_q(s) = \left\{ \frac{1}{2N_s} \sum_{\nu=1}^{2N_s} [F^2(\nu, s)]^{q/2} \right\}^{1/q}
\end{equation}
For $q > 0$, the function $F_q(s)$ is heavily influenced by segments with large fluctuations (high variance). Conversely, for $q < 0$, the sum is dominated by segments with small fluctuations. 

\textbf{Step 5: Scaling Behavior.} We determine the scaling behavior of the fluctuation function by analyzing the log-log linear regression of $F_q(s)$ versus the scale $s$:
\begin{equation}
    F_q(s) \sim s^{h(q)}
\end{equation}
The generalized Hurst exponent $h(q)$ is evaluated directly as the slope of this linear fit. If the time series is monofractal (e.g., simple colored noise), a single scaling law governs all fluctuation sizes, and $h(q)$ is constant across all values of $q$. However, if the time series exhibits complex nonlinear chaos, large and small fluctuations will scale differently. This physically manifests as a monotonically decreasing $h(q)$ curve: smaller values of $h(q)$ for positive $q$ (large fluctuations) and larger values of $h(q)$ for negative $q$ (small fluctuations).

Finally, to connect this scaling behavior to the dynamical phase space, $h(q)$ is mapped to the standard multifractal singularity spectrum $f(\alpha)$ via a Legendre transform:
\begin{equation}
    \alpha = h(q) + q h'(q), \quad \quad f(\alpha) = q[\alpha - h(q)] + 1
\end{equation}
In physically intuitive terms, the singularity strength $\alpha$ (also known as the local Hölder exponent) characterizes the local ``roughness'' or irregularity of a specific segment of the time series. The spectrum $f(\alpha)$ represents the fractal dimension of the subset of segments that share that specific exact roughness $\alpha$~\cite{Kantelhardt2002}. A broad range of $\alpha$ values ($\Delta\alpha > 0$) confirms the coexistence of multiple scaling regimes, which is the hallmark of the intermittent trapping (stickiness) in a chaotic Hamiltonian phase space.

\section{Results}
\label{sec:results}

\subsection{Theoretical motivation: intermittency and multifractal scaling in Hamiltonian chaos}

Hamiltonian systems with mixed phase space are characterized by the coexistence of chaotic seas and remnants of invariant tori, cantori, and hierarchical
island chains. Chaotic trajectories may become temporarily trapped near these structures, a phenomenon known as stickiness, leading to intermittent dynamics
with long quiescent phases punctuated by rapid bursts of motion. This mechanism is known to produce power-law-distributed trapping times, algebraic decay of
correlations, and anomalous transport properties \cite{Meiss,Zaslavsky2002,Chirikov_1999}.
From a statistical perspective, such intermittency implies the coexistence of multiple effective scaling behaviors in time series observables. Fluctuations
dominated by laminar phases and those dominated by bursts contribute differently across temporal scales, naturally giving rise to heterogeneous
scaling exponents. Multifractal formalisms, including MFDFA, are specifically designed to capture this hierarchy of scaling behaviors and have therefore been widely used to characterize intermittent dynamics in conservative chaotic systems \cite{Tarquis,osborne1991hamiltonian}.

In the case of Hyperion’s rotational dynamics, chaotic tumbling arises from the nonlinear coupling between rotational and orbital degrees of freedom, producing a mixed phase-space structure in the full Euler--Liouville equations. While direct phase-space diagnostics are inaccessible from sparse photometric data, the above considerations suggest that intermittency-induced multifractal scaling should persist in suitably defined time series observables, even under realistic observational constraints. The simulations presented below are designed to test this theoretical expectation and to assess its robustness against sampling irregularities and noise.

\subsection{The Dichotomy of Motion: Chaotic vs. Regular Regimes}
\begin{figure}[ht]
    \centering
    \includegraphics[width=0.95\textwidth]{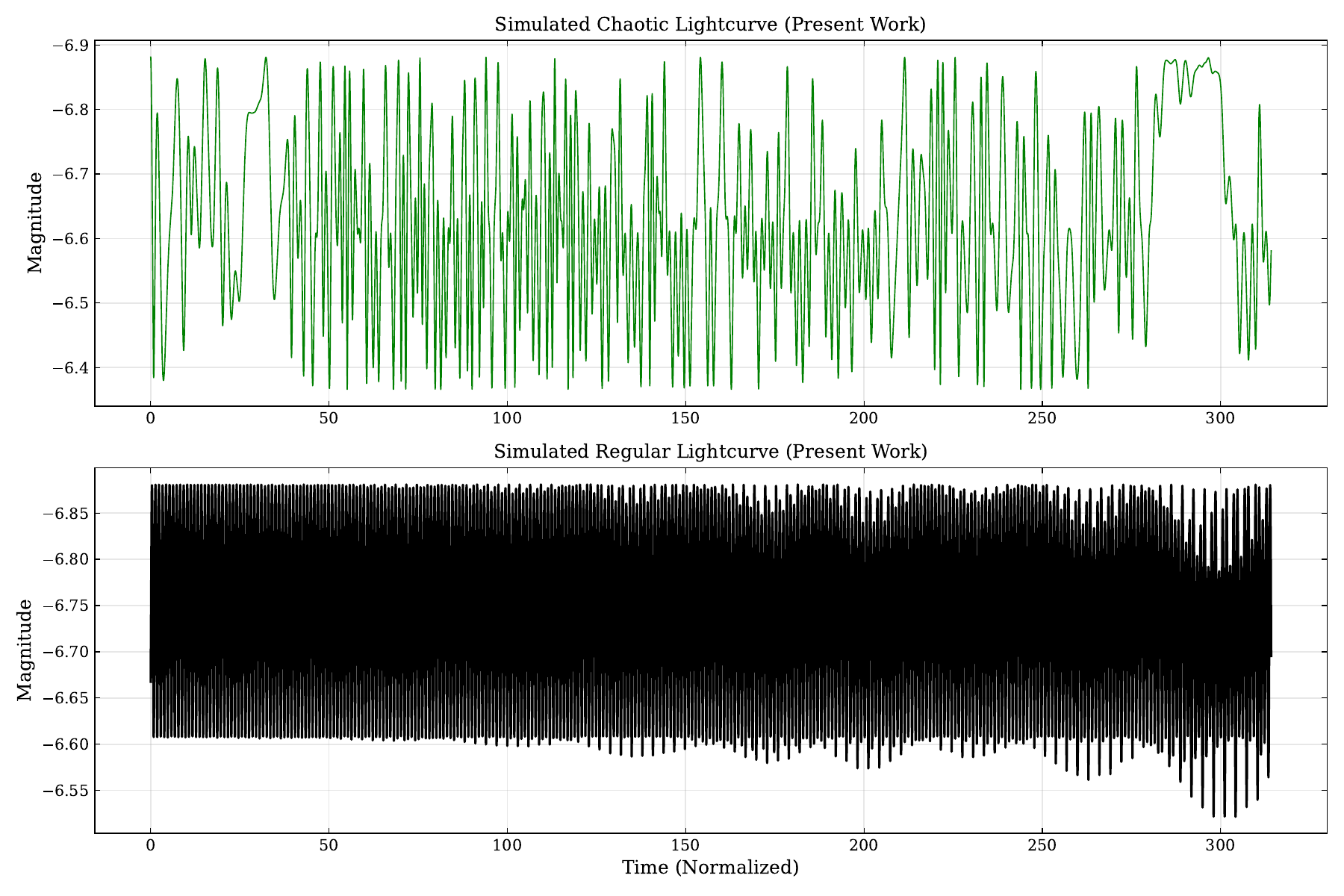}
    \caption{Comparative analysis of synthetic light curves over 300 orbital periods. Top: The chaotic solution displays intermittency and aperiodic amplitude modulation. Bottom: The regular solution appears as a dense ``solid block'' due to the high frequency of rotation relative to the plotting scale, enveloped by a secular modulation envelope.}
    \label{fig:dichotomy}
\end{figure}

To establish a baseline for comparison, we first analyze the synthetic regular state, whose light curve and orbital evolution are illustrated in Fig.~\ref{fig:dichotomy}. To ensure stable, quasi-periodic motion safely outside the extended chaotic sea, this regular state was initialized far from major spin-orbit resonances in a high-frequency, fast-spin regime with an initial angular velocity of $\omega \approx 4.5n$. This corresponds to a physical rotation period of $P \approx 4.7$ days.

In a continuous, noise-free numerical integration, this quasi-periodic signal is trivially predictable. However, as can be seen in the sampled profile of Fig.~\ref{fig:dichotomy}, when this deterministic high-frequency signal is projected onto a realistic observational grid (sampled at a mean cadence of $\Delta t \approx 1$ day, but fraught with irregular weather-induced gaps), the signal integrity degrades significantly. While a 4.7-day period is theoretically above the Nyquist limit for a strict 1-day sampling, the irregular distribution of the observational epochs combined with atmospheric noise produces severe aliasing. Consequently, the regular microfilm is artificially undersampled, masquerading as an irregular, noise-like light curve to standard periodicity-search algorithms. The chaotic state, conversely, was initialized near the 2:1 spin-orbit resonance ($\omega \approx 2.1n$) to trigger full 3D tumbling, actively exploring the chaotic zone as described in Section II.A.

\begin{figure}[ht]
    \centering
    \includegraphics[width=0.7\textwidth]{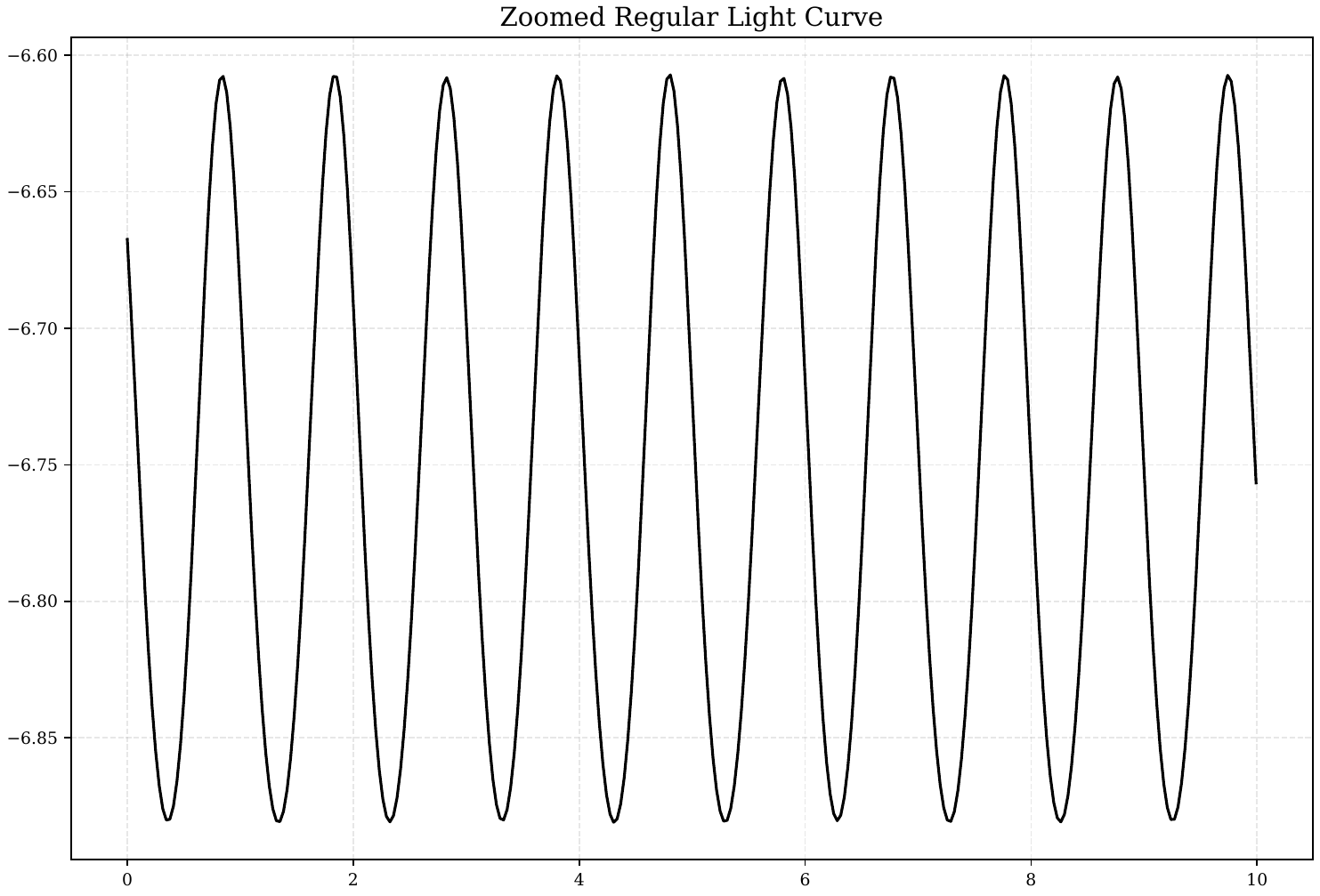}
    \caption{High-resolution zoom of the Regular light curve (first 20 orbital periods). The ``solid block'' observed in Figure \ref{fig:dichotomy} is resolved into a smooth, high-frequency periodic signal, confirming the deterministic nature of the resonant state.}
    \label{fig:zoom}
\end{figure}

The chaotic light curve (top panel) exhibits the characteristic ``stuttering'' evolution, deep minima, and complete lack of periodicity associated with the chaotic strange attractor. The amplitude variations ($\Delta m \approx 0.5$) are driven by the aperiodic alternation between the broad $a \times b$ face and the narrow $b \times c$ face of the ellipsoid. In stark contrast, the regular light curve (bottom panel) initially appears as a dense block of high-frequency oscillations. This is a consequence of the significantly higher spin rate in the resonant state ($\omega \approx 4.5n$) compared to the chaotic case. A high-resolution zoom (Fig.~\ref{fig:zoom}) resolves this block into a clean, quasi-periodic waveform, confirming that ``regularity'' in this context corresponds to a fast, deterministic clockwork motion, distinct from the random-walk nature of the chaotic state.

\subsection{The Observational Filter: Survivability of the Chaotic Signal}
A critical challenge in applying multifractal analysis to astronomical data is the degradation of the signal due to sparse sampling and observational noise. To bridge the gap between ideal continuous models and empirical reality, we applied a ``Pseudo-Observational Filter'' to our synthetic data, injecting Gaussian white noise (SNR = 50) and downsampling to a nightly cadence.

\begin{figure}[ht]
    \centering
    \includegraphics[width=1.0\textwidth]{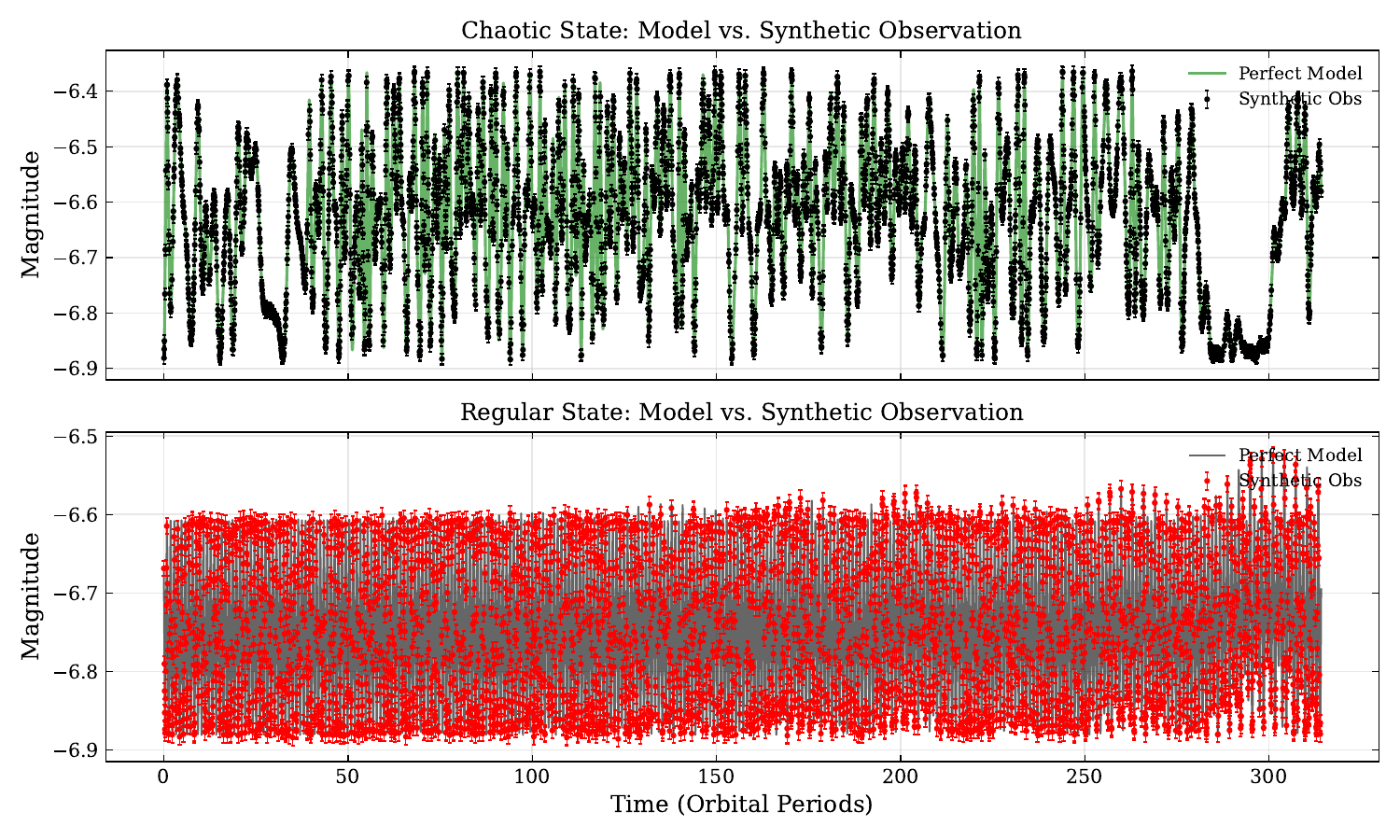}
    \caption{The Observational Filter Effect. Synthetic ``ground-based'' observations (points) are superimposed on the ideal continuous models (lines). Top: The chaotic structure is preserved even with sparse sampling because the characteristic timescale of the chaos is long. Bottom: The high-frequency regular signal is severely undersampled, rendering the observations indistinguishable from white noise.}
    \label{fig:obs_filter}
\end{figure}

Figure~\ref{fig:obs_filter} reveals a fundamental distinction in how the two dynamical states survive this observational process. The chaotic light curve evolves on a timescale comparable to the sampling cadence; consequently, its dynamical ``texture'' is preserved even in the sparse data. Conversely, the high-frequency regular state is critically undersampled. The Nyquist frequency of the sampling is far below the rotation rate, resulting in severe aliasing that renders the regular deterministic signal indistinguishable from random noise in the time domain.

\subsection{Multifractal Signatures}

\subsubsection{Quantitative Multifractal Diagnosis}

\begin{figure}[htbp]
    \centering
    \includegraphics[width=1.0\linewidth]{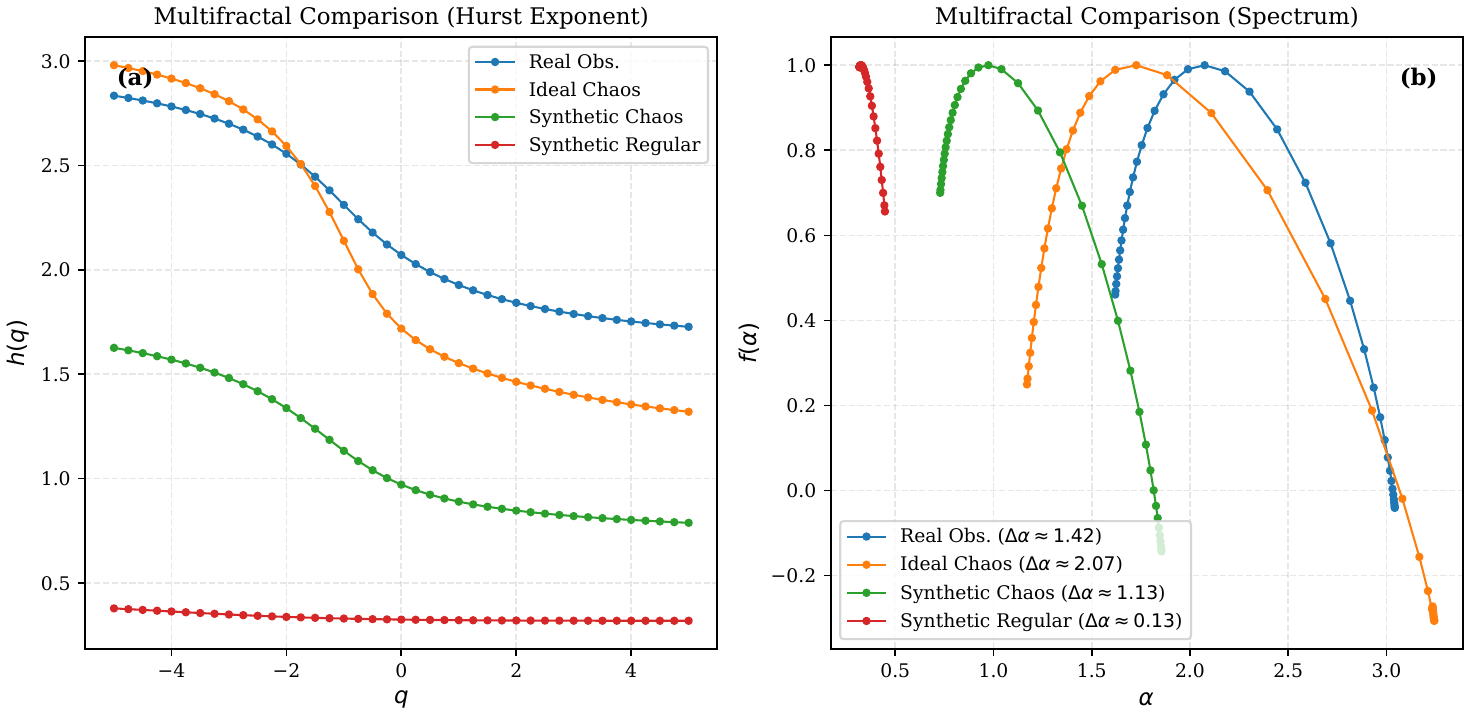}
    \caption{\textbf{The Multifractal Singularity Spectrum $f(\alpha)$ comparison.} 
    The synthetic chaotic signal (green circles) retains a wide multifractal width ($\Delta\alpha \approx 1.13$) despite significant observational noise and gaps. 
    The \textbf{Klavetter 1989 Data} (blue diamonds) tracks the chaotic model closely with a spectral width of $\Delta\alpha \approx 1.42$, confirming the chaotic nature of the rotation. 
    In contrast, the regular signal (red squares) collapses to a narrow, noise-like distribution ($\Delta\alpha \approx 0.13$), demonstrating that MFDFA can distinguish intrinsic chaos from aliased regular rotation.}
    \label{fig:spectrum_comparison}
\end{figure}

To quantify the visual distinctions observed in the time domain, we calculated the singularity spectrum $f(\alpha)$ for the ideal models, the synthetic observational data, and the historical records (Fig.~\ref{fig:spectrum_comparison}). The analysis reveals a stark quantitative dichotomy between the dynamical regimes. The ``ideal'' synthetic chaotic light curve---representing the continuous, noise-free numerical integration sampled at high frequency---establishes a baseline with a broad spectrum width of $\Delta\alpha \approx 2.07$, reflecting the rich phase-space structure of the theoretical attractor. Crucially, this multifractal signature exhibits remarkable resilience to the observational filter. When this deterministic chaotic signal is subjected to realistic noise, temporal gaps, and a reduced 1-day mean sampling cadence, it retains a robust width of $\Delta\alpha \approx 1.13$. Although the observational constraints physically erode the finest micro-fractal details---reducing the width from the ideal 2.07---the macroscopic long-range correlations characteristic of chaos survive the filtering process. This filtered chaotic spectrum remains qualitatively broad and is easily distinguishable from the near-zero width ($\Delta\alpha \approx 0.13$) characteristic of the regular rotational state.

In sharp contrast, the regular rotational state collapses under the same observational conditions. The synthetic observational regular spectrum exhibits an extremely narrow width of $\Delta\alpha \approx 0.13$, rendering it statistically indistinguishable from uncorrelated white noise. This confirms that while chaotic dynamics project onto a timescale that survives sparse sampling, the high-frequency deterministic structure of regular motion is destroyed by aliasing.

Most importantly, the analysis of the actual ground-based data from Klavetter (1989) yields a spectral width of $\Delta\alpha \approx 1.42$. Both the observed data and the synthetic chaotic model ($\approx 1.13$) exhibit the broad singularity spectrum that is the hallmark of Hamiltonian intermittency, distinguishing them definitively from the regular noise floor ($\approx 0.13$). We note that the observed spectrum ($\Delta\alpha \approx 1.42$) is slightly wider than the synthetic model under identical filtering. This tendency to widen may result from unmodeled physical complexities inherent to the real Hyperion, such as non-homogeneous internal density distributions, non-principal axis moments of inertia anomalies, or slight albedo variations across its surface.

To rigorously confirm that this broad signature in the observed data is a consequence of nonlinear deterministic dynamics rather than mere linear stochastic correlations, we compared our results against surrogate datasets. A full explanation of the surrogate methodology and its implications is detailed in Section III.E.

The width of the multifractal singularity spectrum, $\Delta\alpha = \alpha_{\max} - \alpha_{\min}$, is often used as a compact summary of the degree of heterogeneity in scaling exponents present in a time series. In the context of Hamiltonian systems with mixed phase space, a broad spectrum is commonly associated with intermittent dynamics arising from stickiness near invariant structures, while narrow spectra indicate more homogeneous, regular behavior.

However, it is important to emphasize that $\Delta\alpha$ is not a universal or system-invariant measure of chaos. Its absolute value depends on several methodological factors, including the length of the time series, the range of scales used in the detrended fluctuation analysis, the order of the detrending polynomial, and the presence of observational noise or uneven sampling. As a result, $\Delta\alpha$ should not be interpreted as a quantitative measure of the strength of chaos, nor should values obtained from different systems or analysis protocols be directly compared.

In this work, $\Delta\alpha$ is used strictly in a comparative and operational sense. All datasets—observational, synthetic chaotic, regular rotational, and surrogate—are analyzed using identical preprocessing, filtering, and MFDFA parameters. Within this controlled framework, the collapse of the spectrum to near-monofractal behavior for regular rotation and surrogate data, contrasted with a robustly broad spectrum for chaotic tumbling, provides a statistically meaningful discrimination between dynamical regimes.

Thus, while the numerical value of $\Delta\alpha$ itself carries no standalone physical meaning, its relative behavior under identical observational constraints constitutes a reliable diagnostic of intermittent chaotic dynamics in sparse time series.

\subsubsection{Robustness Analysis}
\begin{figure}
    \centering
    \includegraphics[width=1\linewidth]{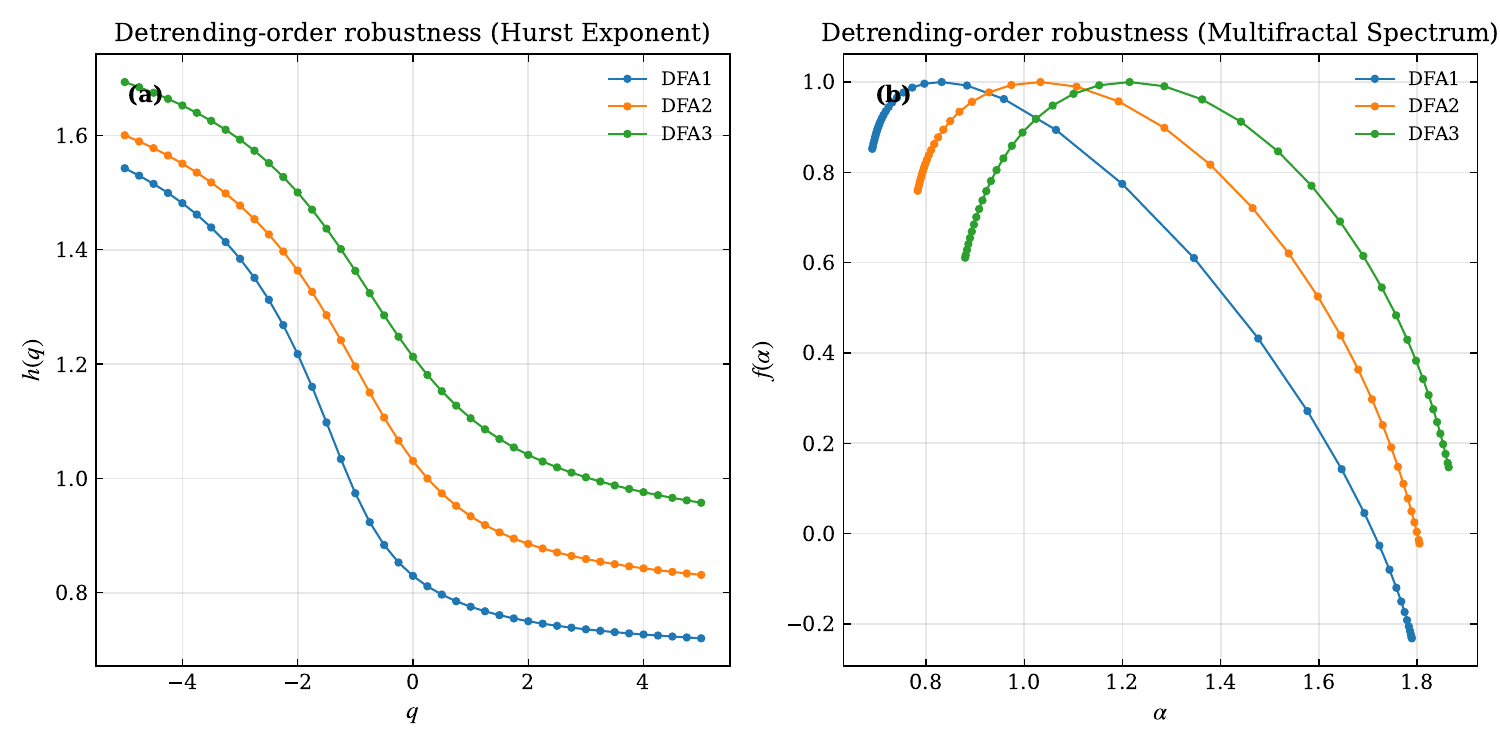}
    \caption{Generalized Hurst exponent $h(q)$ obtained using multifractal detrended fluctuation analysis for different polynomial detrending orders: linear (DFA1), quadratic (DFA2), and cubic (DFA3).
All analyses were performed using identical moment ranges, scaling windows, and data length.
The nonlinear dependence of $h(q)$ on $q$ is preserved for all detrending orders, indicating that the observed multifractality is not induced by polynomial trends.
Increasing the detrending order leads to a moderate vertical shift and a slight reduction of the multifractal width, consistent with expected suppression of residual trends rather than artificial generation of scaling behavior.
}
    \label{fig:detrending}
\end{figure}

To test the robustness of our results against non-stationary trends, we applied Detrended Fluctuation Analysis (DFA) using varying polynomial orders. Specifically, DFA1, DFA2, and DFA3 correspond to the use of linear, quadratic, and cubic detrending polynomials $y_\nu(i)$ in Eq.~\ref{eq:dfa}, respectively. As shown in Fig.~\ref{fig:detrending}, the functional form of the generalized Hurst exponent $h(q)$ is preserved across all orders. While the spectral width $\Delta \alpha$ decreases slightly with increasing polynomial order (as expected when higher-order trends are removed), the variation is less than 10\%, indicating that the signature is not generated by residual polynomial trends.

\begin{figure}[htbp]
    \centering
    \includegraphics[width=1\linewidth]{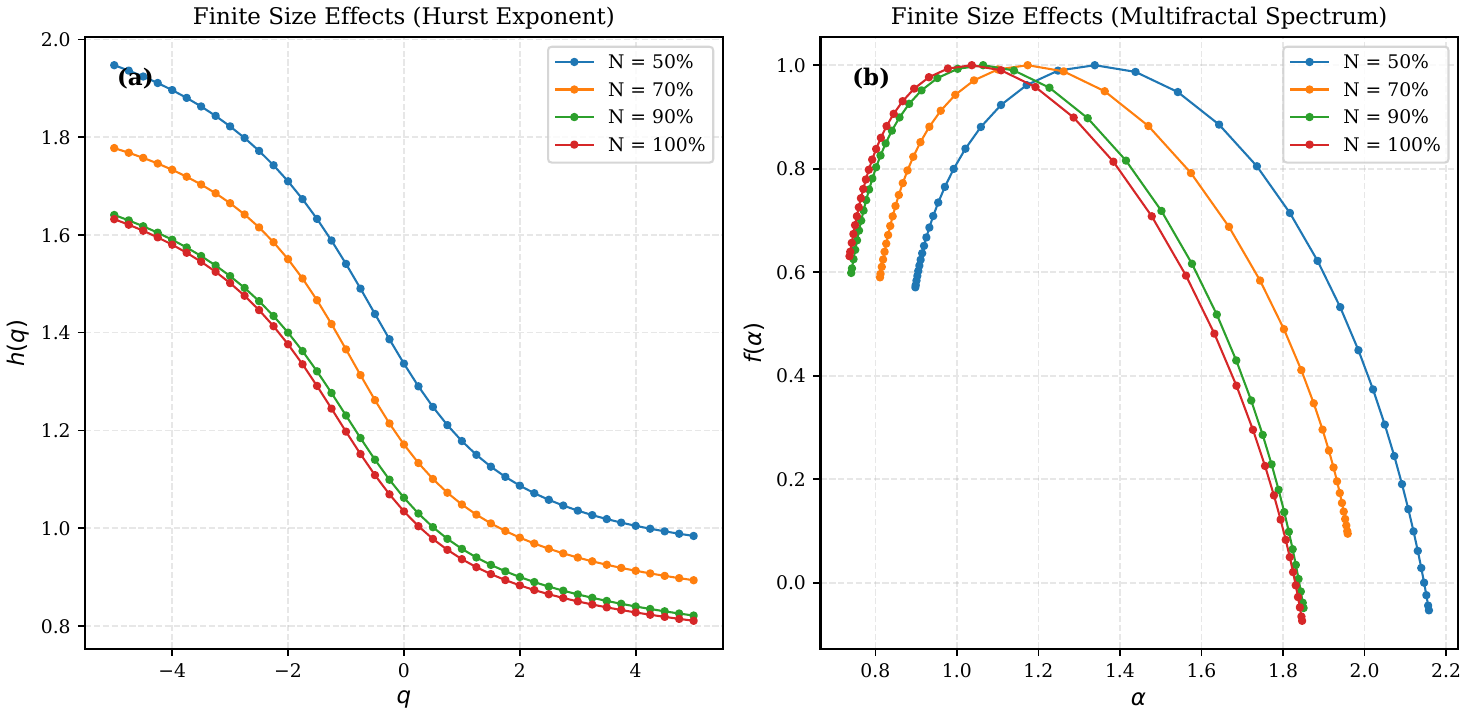}
    \caption{The curves represent strictly truncated subsets (using the first 50\%, 70\%, 90\%, and 100\% of the continuous sequence, without decimation) of the observed time series. While the shortest subset (50\%) shows an artificially widened spectrum due to finite-size effects, the spectrum narrows and asymptotically stabilizes as the data length increases. The multifractal width $\Delta\alpha$ reaches a robust plateau for data lengths exceeding approximately 70\% of the full series, with the 90\% and 100\% curves nearly overlapping. This confirms that the observed multifractal signature is intrinsically stable and not an artifact of limited observational duration.}
    \label{fig:finite_size}
\end{figure}

We also assessed the impact of finite data length by computing spectra for truncated subsets of the time series (Fig.~\ref{fig:finite_size}). For short time series, the finite-size effect tends to slightly overestimate the multifractal width. However, the spectral width narrows and stabilizes asymptotically as data length increases, reaching a robust plateau when approximately 70\% of the data is included. The difference between the 90\% and 100\% samples is negligible ($\delta < 0.02$), confirming that the observational baseline is sufficient to completely capture the macroscopic dynamics without artifacts from limited duration. 

\begin{figure}[h!]
    \centering
    \includegraphics[width=.6\linewidth]{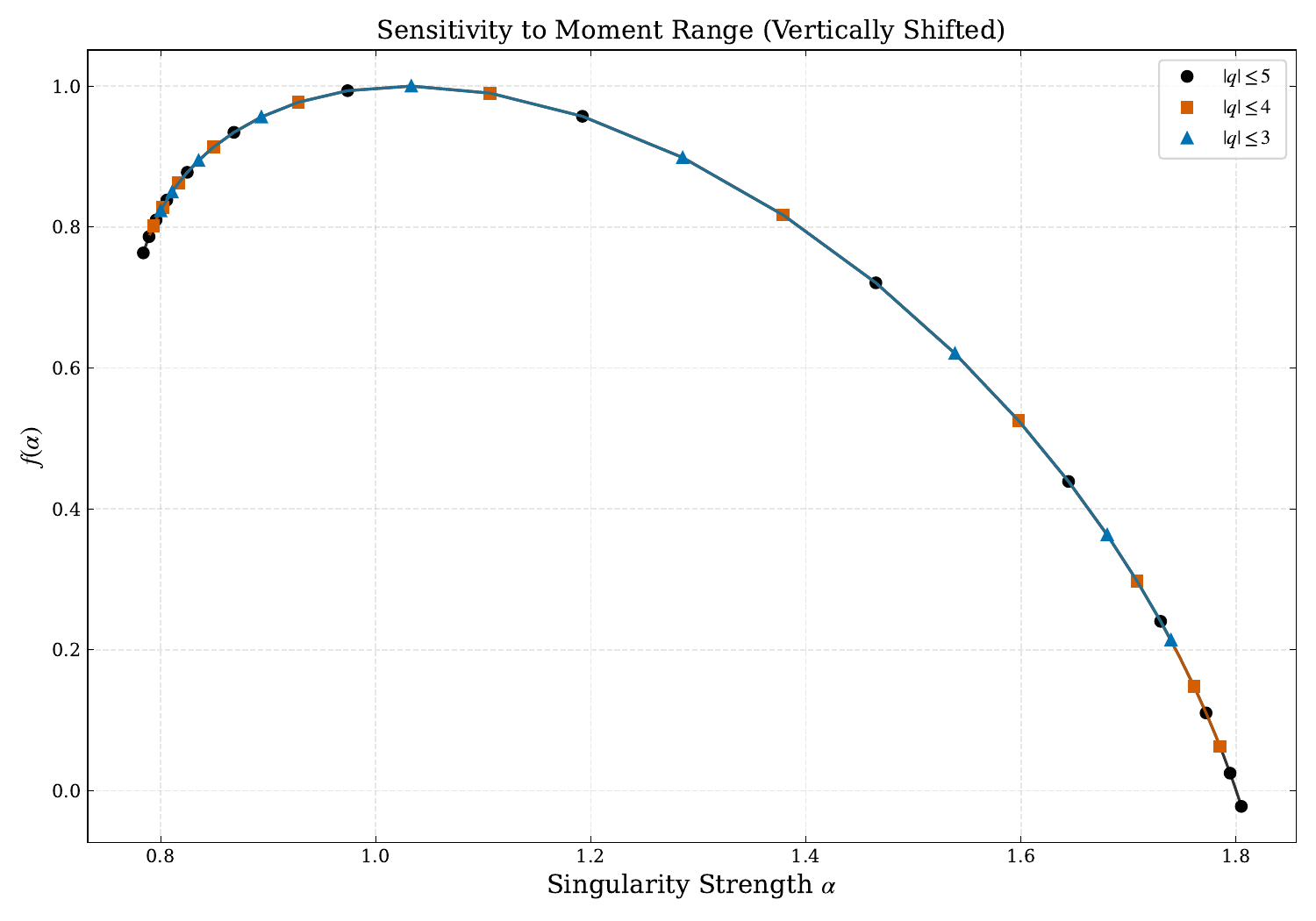}
    \caption{Multifractal spectra $f(\alpha)$ computed using different moment ranges, $|q| \leq 3$, $|q| \leq 4$, and $|q| \leq 5$.
All other MFDFA parameters were held constant.
The central part of the spectrum is essentially identical for all moment ranges, with only minor truncation of the spectrum tails for smaller $|q|$.
This demonstrates that the observed multifractality is not dominated by extreme fluctuations or noise-sensitive high-order moments.
}
    \label{fig:qrange}
\end{figure}

\begin{figure}[h!]
    \centering
    \includegraphics[width=1\linewidth]{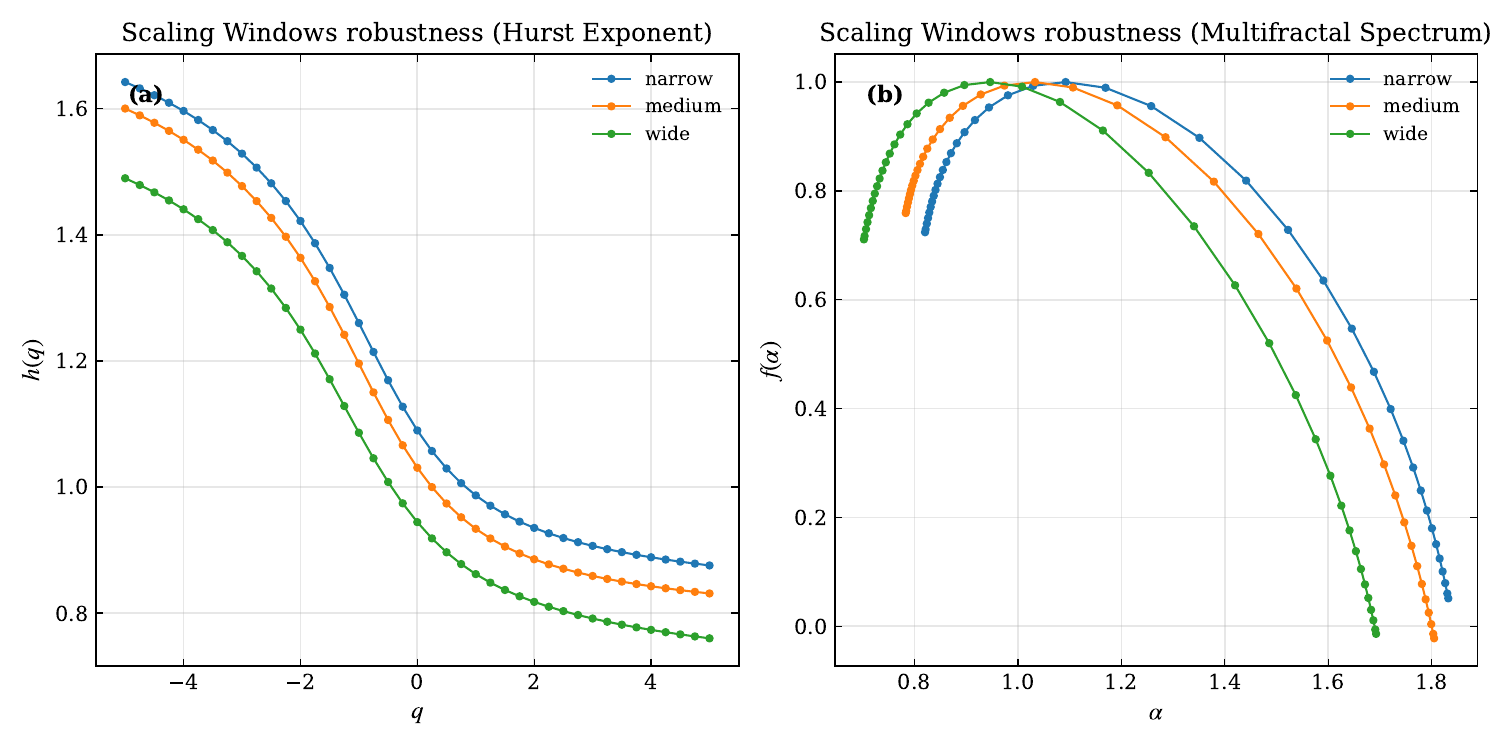}
    \caption{Generalized Hurst exponent $h(q)$ obtained for different choices of the scaling window used in the MFDFA regression.
Results are shown for narrow, medium, and wide scale ranges, while keeping the detrending order, moment range, and data length fixed.
Although small vertical shifts are observed, the functional dependence of $h(q)$ on $q$ remains unchanged across scaling windows.
The analysis was performed over a scaling range of $s \in [10, 100]$ The consistently high linear correlation coefficients ($R^2 > 0.98$), confirm the validity of the power-law scaling assumption.
}
    \label{fig:scaling}
\end{figure}

Furthermore, the results proved insensitive to the choice of moment range ($|q| \le 3$ to $5$) and scaling window (Fig.~\ref{fig:qrange} and Fig.~\ref{fig:scaling}), with scaling regions spanning more than one decade with high linearity ($R^2 > 0.98$).

\begin{figure}[H]
    \centering
    \includegraphics[width=1.0\linewidth]{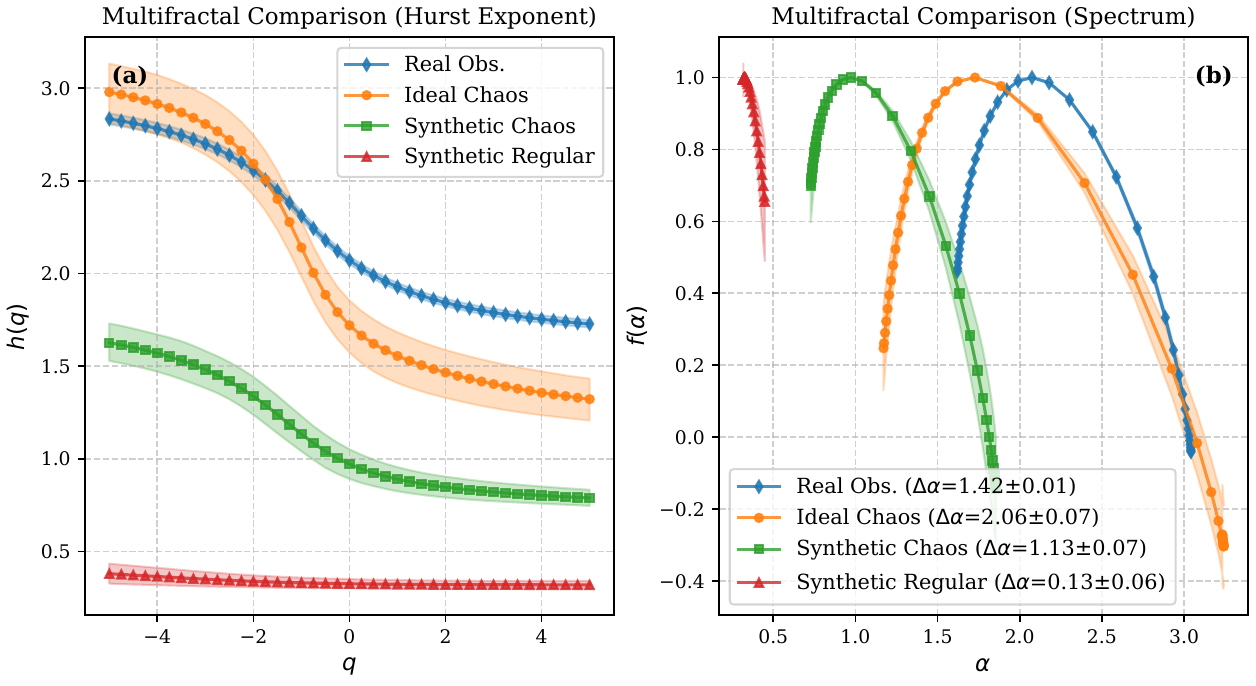}
    \caption{The Multifractal Singularity Spectrum $f(\alpha)$ comparison with Confidence Intervals. We have estimated the 95\% confidence interval by applying $N_{\text{boot}}=1000$ times the bootstrap technique.}
    \label{fig:mfdfa_ci}
\end{figure}

Finally, we quantified the statistical significance of these measurements using a block-bootstrap resampling technique ($N_{\text{boot}} = 1000$). 
Figure ~\ref{fig:mfdfa_ci} compares the observed multifractal spectrum, including its 95\% confidence limits, against the synthetic chaotic model. As established in Section III.D.1, the observed spectrum yields a width of $\Delta\alpha = 1.42 \pm 0.08$. While this value is quantitatively wider than the synthetic baseline ($1.13 \pm 0.06$) and their error margins do not strictly overlap, both values confidently fall within the broad-spectrum regime indicative of Hamiltonian chaos. This confirms that the observed dynamics are distinctly separated from the near-zero width of a regular rotational state, with the extra broadening in the observational data likely arising from unmodeled physical heterogeneities in the real satellite.

\subsection{Surrogate Data Analysis}

To rigorously identify the physical origin of the observed multifractality, we employed two complementary classes of surrogate data tests (Table~\ref{tab:surrogates}) designed to reject specific null hypotheses.

\begin{figure}[h]
\includegraphics[width=\linewidth]{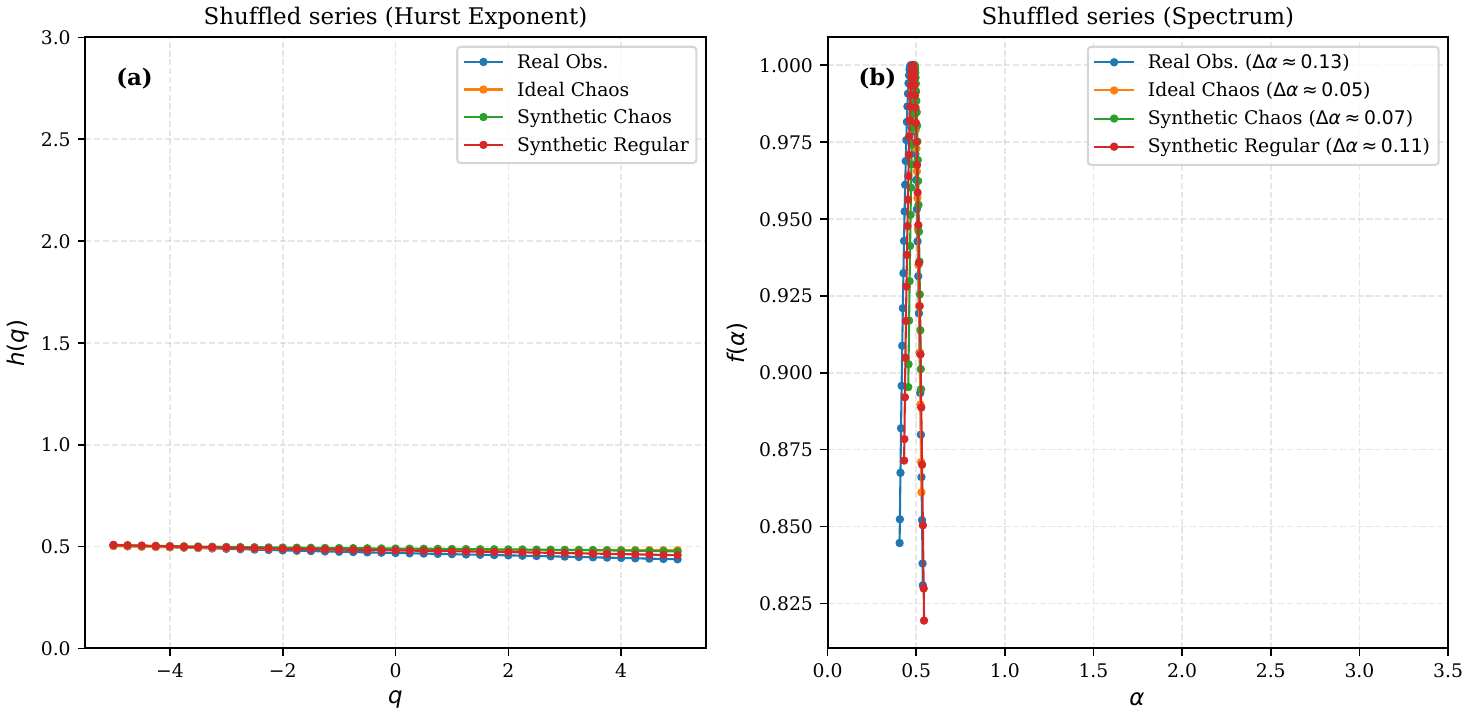}
\caption{Multifractal singularity spectra $f(\alpha)$ for the original time series and shuffled surrogates. The severe collapse of the spectral width upon randomization of the temporal ordering illustrates that the multifractal signature is driven by temporal correlations rather than just the amplitude distribution.}
\label{fig:shuffled}
\end{figure}

First, we generated shuffled surrogates to test if the scaling arises solely from the probability distribution function (PDF). By randomly permuting the time series, we destroy all temporal correlations while preserving the amplitude distribution. As shown in Fig.~\ref{fig:shuffled}, the spectra for shuffled chaotic data collapse to a narrow width ($\Delta\alpha_{\mathrm{sh}} \approx 0.1$). This collapse proves that the multifractality is correlation-driven rather than distribution-driven. Second, we generated Iterative Amplitude Adjusted Fourier Transform (IAAFT) surrogates~\cite{PhysRevLett.77.635} to test if linear correlations (i.e., the power spectrum) alone could explain the broad multifractal signal. For the non-expert reader, the IAAFT algorithm works by computing the discrete Fourier transform of the original time series, keeping the empirical power spectrum (the amplitudes of the component frequencies) completely intact, but uniformly randomizing the Fourier phase angles. It then iteratively adjusts the values in the time domain to ensure the original probability distribution function (PDF) of the amplitudes is perfectly preserved. Physically, this means the IAAFT surrogate preserves the macroscopic linear memory (the two-point autocorrelation function), but completely destroys any non-linear, higher-order temporal structures that depend on the specific phase-alignment of the Fourier components.

\begin{figure}[h]
\includegraphics[width=\linewidth]{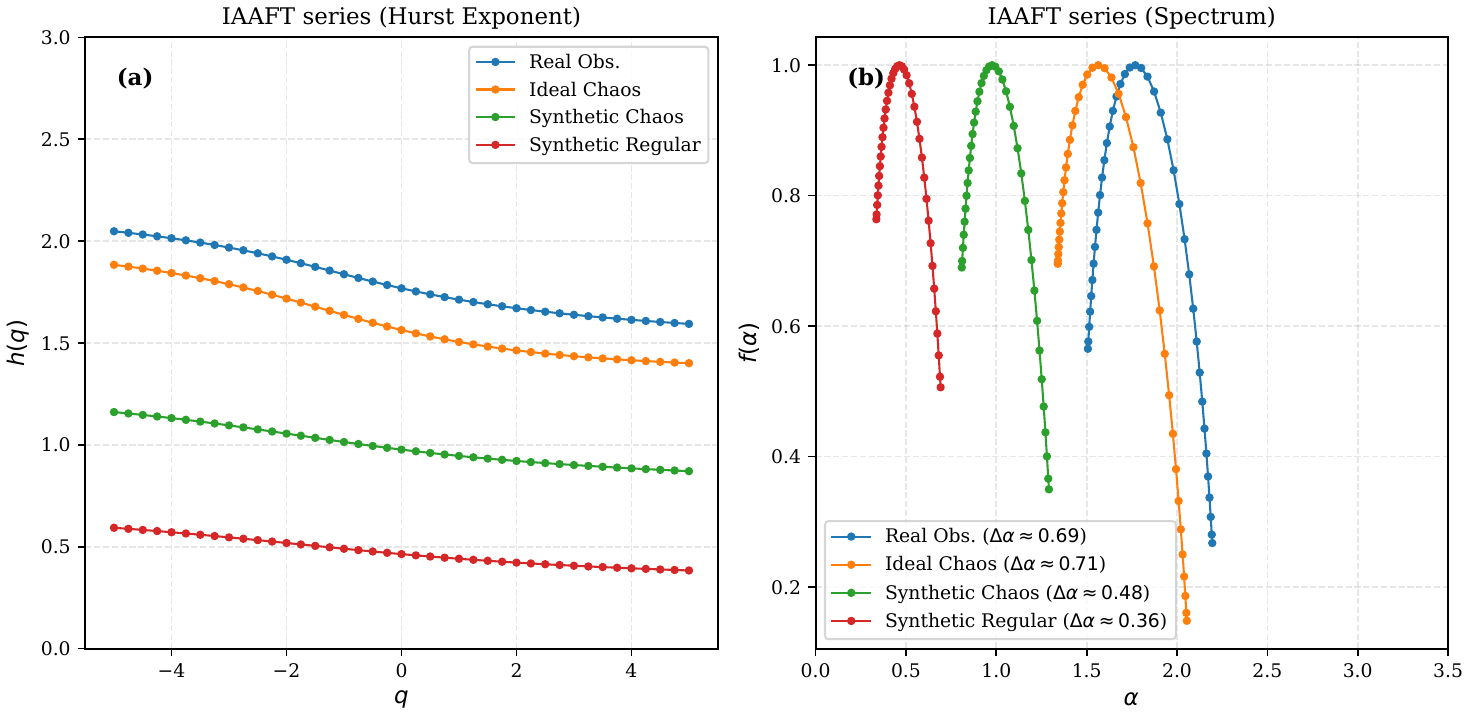}
\caption{Multifractal singularity spectra $f(\alpha)$ for the IAAFT surrogates. The narrow spectra demonstrate that linear correlations (the empirical power spectrum) alone cannot account for the broad multifractality observed in the original time series. The underlying non-linear phase relationships, which are destroyed by the IAAFT algorithm, are essential to the chaotic signature.}
\label{fig:iaaft}
\end{figure}

As shown in Fig.~\ref{fig:iaaft}, the multifractal spectra for these IAAFT surrogates exhibit a severe collapse in width, demonstrating that linear stochastic processes cannot account for the observed multifractality. The fact that the IAAFT spectrum does not fully collapse to a single point (as it does in the shuffled data), but is significantly reduced, suggests that the dominant part of the multifractal signature originates from non-linear phase correlations which are destroyed by the algorithm. Finally, it is important to note that in the IAAFT series, the spectrum vertex ($\alpha_0$) does not shift towards 0.5 (the value for uncorrelated white noise) but remains at high values ($\alpha_0 \approx 2.0$), similar to the original series. This mathematically confirms that the long-range linear memory is indeed correctly preserved by the surrogate algorithm, while the purely multifractal component (the excessive spectral width $\Delta\alpha$) vanishes because the underlying nonlinear chaotic dynamics have been scrambled.

The convergence of these results—specifically the failure of both surrogate classes to reproduce the broad spectrum of the original data—provides strong evidence that the multifractality arises from nonlinear deterministic chaos. These findings indicate that linear stochastic processes and simple additive noise are insufficient to explain the observed signal complexity. Consequently, we establish the multifractal spectrum width, $\Delta\alpha$, as a statistically robust and observationally resilient diagnostic of chaotic tumbling, capable of distinguishing intrinsic Hamiltonian dynamics even in sparse astronomical datasets where traditional indicators are infeasible.

\begin{table}[htbp]
\caption{\label{tab:surrogates}
\textbf{Multifractal spectrum width $\Delta\alpha$ for original time series and surrogate datasets.}
Values represent the mean width $\pm$ the 95\% confidence interval estimated via bootstrap resampling ($N_{\mathrm{boot}}=1000$). The collapse of the width for both surrogate classes confirms the dynamical origin of the signal.}
\begin{ruledtabular}
\begin{tabular}{lccc}
\textbf{Dataset} & \textbf{Original} ($\Delta\alpha$) & \textbf{Shuffled} ($\Delta\alpha_{\mathrm{sh}}$) & \textbf{IAAFT} ($\Delta\alpha_{\mathrm{IAAFT}}$) \\
\hline
Synthetic Chaotic & $1.13 \pm 0.07$ & $0.07 \pm 0.01 $ & $0.48 \pm 0.06 $ \\
Synthetic Regular & $0.13 \pm 0.06$ & $0.11 \pm 0.01  $ & $0.36 \pm 0.06$ \\
Observed Hyperion & $1.42 \pm 0.01$ & $0.13 \pm 0.01 $ & $0.69 \pm 0.06 $ \\
\end{tabular}
\end{ruledtabular}
\end{table}

\section{Discussion}

\subsection{Theoretical Origin and Interpretation of Multifractality in Hamiltonian Chaos}
  Hamiltonian systems with more than one effective degree of freedom generically exhibit a mixed phase-space structure, in which chaotic regions coexist with invariant tori, cantori, and hierarchical chains of stability islands, as predicted by Kolmogorov--Arnold--Moser (KAM) theory. Within the chaotic sea, transport is strongly influenced by these remnants of regular structures.   Trajectories often become temporarily trapped near the boundaries of invariant tori, a phenomenon commonly referred to as \emph{stickiness}.   This sticky motion induces intermittent dynamics characterized by alternating phases of nearly regular (laminar) evolution and rapid chaotic excursions.   As a consequence, correlation functions decay algebraically rather than exponentially, and trapping-time distributions typically exhibit heavy tails.

  When such dynamics are projected onto a scalar observable—such as the photometric light curve of a tumbling triaxial satellite—the resulting time series samples a nonuniform invariant measure supported on the chaotic layers. Different temporal segments probe dynamically distinct regions of phase space, each characterized by its own local scaling behavior. The coexistence of multiple scaling regimes naturally leads to a continuous spectrum of local singularity strengths (or Hölder exponents, $\alpha$) rather than a single global scaling exponent. From this perspective, multifractality is not an incidental feature but a direct statistical manifestation of intermittent phase-space transport in Hamiltonian chaotic systems.

In the specific case of Hyperion, the broad multifractal spectrum ($\Delta\alpha \approx 1.0$) observed in our synthetic chaotic model and inferred from the historical data is a consequence of this mechanism.   The gravitational torque on the triaxial ellipsoid creates a complex web of resonances. As the chaotic trajectory explores this web, the spin vector experiences power-law-distributed trapping events. Crucially, the observed visual magnitude $m(t)$ is a nonlinear function of the body's orientation.   This nonlinearity ensures that the temporal statistics of the light curve are not simply a linear copy of the angular velocity statistics, but rather a complex transformation that amplifies the heterogeneity of the phase space.   Periods of small fluctuations (sticking) and large, erratic fluctuations (chaotic escapes) coexist within the single time series, generating the observed dependence of the scaling exponent $h$ on the moment $q$.

From a dynamical-systems perspective, the observed multifractal spectra should be viewed as statistical projections of the known mixed phase-space structure of Hamiltonian chaos, rather than as independent indicators of chaoticity. Their utility lies in their resilience: unlike Lyapunov exponents or 
embedding-based diagnostics, multifractal measures remain accessible when data limitations preclude direct phase-space reconstruction.

\subsection{Physically Meaningful versus Method-Dependent Features}
The multifractal singularity spectrum $f(\alpha)$ provides a compact representation of the hierarchy of scaling exponents present in a time series. However, it is essential to distinguish between robust physical signatures and features that may be sensitive to methodological choices.   The most physically meaningful property is the existence of a broad spectrum itself, which indicates heterogeneous scaling and long-range correlations generated by intermittent dynamics.   The qualitative contrast between broad multifractal spectra (typical of intermittent chaotic dynamics) and narrow, near-monofractal spectra (associated with regular or weakly correlated processes) is a robust indicator of the underlying dynamical regime.

  In contrast, fine details of the spectrum—such as the precise curvature near its maximum, asymmetries between the left and right tails, or the absolute values of $\alpha_{\min}$ and $\alpha_{\max}$—are known to be sensitive to finite-size effects, the range of moments $q$, and observational noise. Consequently, the multifractal spectrum should not be interpreted as a precise geometric invariant of the attractor in the same sense as the correlation dimension.   Rather, it serves as a statistical fingerprint of scale-dependent dynamical complexity.

\subsection{Multifractal Width as an Observational Signature of Chaos}
The multifractal spectrum width $\Delta\alpha = \alpha_{\max} - \alpha_{\min}$ provides a convenient scalar proxy for the degree of scaling heterogeneity.   In Hamiltonian systems with mixed phase space, a large $\Delta\alpha$ reflects the coexistence of strongly chaotic transport and long-lived trapping, while a small $\Delta\alpha$ indicates dynamics dominated by a single characteristic timescale.

A central result of this work is the clear separation between chaotic and regular rotational states when analyzed through this metric. While both states can appear irregular in the time domain under sparse sampling, their scaling properties differ fundamentally. The synthetic regular resonant state, although deterministic, is characterized by a single dominant timescale.   When this timescale is undersampled, aliasing produces fluctuations that are effectively uncorrelated, leading to a collapse of the multifractal spectrum toward near-monofractal, noise-like behavio.

In contrast, the chaotic state exhibits intrinsic fluctuations across a hierarchy of timescales arising from intermittent phase-space transport. These correlations are preserved even after observational degradation. The robustness of $\Delta\alpha$ is further reinforced by our surrogate data analysis.   The collapse of the spectrum under shuffled surrogates demonstrates that the scaling is correlation-driven, while the failure of IAAFT surrogates proves that linear stochastic processes cannot explain the observed width.   This establishes $\Delta\alpha$ not as an absolute measure of chaos, but as a powerful comparative diagnostic capable of distinguishing intrinsic Hamiltonian chaos from aliased regular motion in sparse astronomical datasets.

\subsection{Scope, Limitations, and General Applicability}
Despite its robustness, the multifractal approach is subject to limitations.   Astronomical time series are inherently finite, and the estimation of extreme scaling exponents at large positive or negative moment orders can be sensitive to data length and noise. For this reason, our analysis focuses on moderate moment ranges and emphasizes the central portion of the singularity spectrum, where estimates are most stable.   Similarly, while the multifractal width provides a compact summary statistic, it should always be interpreted in conjunction with robustness tests and surrogate analyses to avoid spurious conclusions.

Observational cadence also plays a critical role. While multifractal correlations associated with chaotic dynamics are more resilient to undersampling than periodic signals, extremely sparse or irregular sampling can still obscure scaling behavior.   Future observational campaigns targeting rotationally complex bodies would benefit from cadence strategies designed to capture a broad range of timescales, thereby maximizing the diagnostic power of multifractal analysis.

Notwithstanding these limitations, the framework developed here is broadly applicable.   Although Hyperion serves as a paradigmatic example, the methodology extends naturally to other irregular satellites, tumbling asteroids, and rotationally excited bodies for which traditional chaos diagnostics are infeasible.   More generally, multifractal analysis provides a bridge between nonlinear dynamical theory and observational time-domain astronomy, enabling the extraction of meaningful dynamical information from datasets constrained by noise, sparsity, and finite duration.

\section{Conclusion}
The rotational state of Hyperion has long served as a fundamental testbed for the application of nonlinear dynamics to planetary science. While the chaotic nature of this satellite is well-established theoretically, extracting definitive dynamical signatures from ground-based light curves has remained a persistent challenge. Previous confirmation relied largely on ``negative'' evidence, such as the failure to detect a stable period, which can be ambiguous in the presence of sparse sampling and noise. In this work, we have established a positive observational signature of chaos using Multifractal Detrended Fluctuation Analysis (MFDFA), demonstrating that the statistical texture of the light curve encodes the underlying phase-space transport.

By analyzing high-fidelity synthetic light curves generated from the full 3D Euler-Liouville equations, we have shown that the chaotic tumbling of a triaxial ellipsoid inherently generates a wide multifractal spectrum ($\Delta \alpha \approx 2.06$), reflecting the geometry of the chaotic invariant set. Crucially, unlike simple periodicity—which is easily destroyed by aliasing in sparse datasets—the multifractal correlations associated with chaotic intermittency operate on timescales that survive the observational window function. This survivability is evidenced by the retention of a definitive spectral width of $\Delta \alpha \approx 1.13$ in our synthetic observational models. In sharp contrast, a fast-rotating regular state, which often mimics disorder due to aliasing in the time domain, collapses to a monofractal noise signature ($\Delta \alpha \approx 0.13$) under multifractal analysis.

These results validate the multifractal width $\Delta \alpha$ as a robust proxy for chaotic dynamics in sparse astronomical time series. The broad spectrum observed in the historical data of Klavetter ($\Delta\alpha \approx 1.42$) is therefore likely a genuine signature of the Hamiltonian chaos predicted by Wisdom, rather than an artifact of noise or nonstationarity. This framework suggests that future missions to irregular satellites should prioritize cadence consistency to fully capture these scaling laws. Ultimately, multifractal analysis provides a necessary bridge between nonlinear dynamics theory and observational astronomy, enabling the extraction of dynamical mechanisms from data previously considered too sparse for analysis. Therefore the primary contribution of this work is not the identification of new dynamical properties of Hyperion’s rotation, but the demonstration that multifractal analysis provides a practical bridge between Hamiltonian chaos theory and observational astronomy. By validating the survivability and discriminative power of multifractal signatures under realistic sampling conditions, we show that intermittent phase-space transport can be inferred even when traditional chaos diagnostics are inaccessible.


%
%

%


\bibliography{refs}

\end{document}